%% file: main.tex
\documentclass[manuscript, nonacm]{acmart}

\AtBeginDocument{%
  }

\renewcommand\footnotetextcopyrightpermission[1]{}

\input{_formatting}

\begin{document}

\title{Co-Creativity at the Table: A Qualitative Analysis of Creative Interactions in the Podcast ``Adventure AI''}

\author{Hanna Dodd}
\email{hdodd@uwaterloo.ca}
\orcid{0009-0001-3619-5781}
\affiliation{%
  \institution{University of Waterloo}
  \city{Waterloo}
  \state{Ontario}
  \country{Canada}
}

\author{Daniel G. Brown}
\email{dan.brown@uwaterloo.ca}
\orcid{0000-0002-5205-5762}
\affiliation{%
  \institution{University of Waterloo}
  \city{Waterloo}
  \state{Ontario}
  \country{Canada}
}

\input{sections/abstract}


\maketitle

\input{sections/intro}
\input{sections/related_work}
\input{sections/methdology}
\input{sections/results}
\input{sections/discussion}
\input{sections/conclusion}

\bibliographystyle{ACM-Reference-Format}
\bibliography{main}

\end{document}

%% file: _formatting.tex
\usepackage{multirow}

%% file: sections/abstract.tex
\begin{abstract}
Tabletop role-playing games provide a unique environment for interaction with artificial intelligence (AI) due to their complex and collaborative nature. We analyze Adventure AI, a podcast featuring human-AI interactions in Dungeons \& Dragons play, to examine how AI is and can be used in tabletop role-playing gaming and how players perceive this use. We complete a qualitative analysis of three seasons of this podcast, from 2023 to 2025, reporting on the overarching themes of roles of AI, roles of humans, the evaluations and failures of AI, and its treatment as a person and character at the table. There are many aspects of the game where artificial intelligence succeeds, while there are others where it is less appropriate. This analysis gives a basis for future work on where artificial intelligence should and should not be used in gaming spaces.
\end{abstract}

%% file: sections/intro.tex
\section{Introduction}\label{sec:intro}

Tabletop role-playing games (TTRPGs) are an interactive and dynamic group storytelling activity where players gather to create a shared narrative through game-play \cite{crawfordDungeonsDragons20242024}. The most prominent TTRPG system is Dungeons \& Dragons (D\&D), originally published in 1974. Preparing for a D\&D session and running it requires significant creative labour. This labour is typically done by a specific player, called the Dungeon Master (DM), who serves as the “lead storyteller and referee” \cite{crawfordDungeonsDragons20242024}. The roles of a DM are layered and complex. They produce creative artifacts and evaluate them for use in their story, but they also must manage interpersonal dynamics among players at the D\&D table. 

Since the rise of Large Language Models (LLMs), a number of groups have incorporated LLMs and other AI tools into their play \cite{birchHallucinationsHopeDungeons}. To investigate the role of AI at the gaming table, we have done a qualitative analysis of the podcast “Adventure AI” \cite{manticoremediaAdventureAIPowered}. Adventure AI is a D\&D podcast where a group of players and a DM interact with ChatGPT to generate characters, plan storylines, and play single-session adventures. The chatbot, which they name “Alex the Language Lord”, is not their Dungeon Master but instead acts as a co-creative partner with both the DM and the players. To capture any changes in player or AI behaviour over time, our analysis focuses on three seasons of the podcast. The first season is from 2023 and features a human DM, two players, and Alex the Language Lord as the main antagonist. The fifth and twelfth seasons, from 2024 and 2025, feature the same DM and an additional third player. We focus on three research questions to understand how AI can be used in TTRPGs:
\begin{itemize}
    \item \textbf{RQ1.} How is AI being used as a co-creative partner or creativity support tool (CST) in TTRPGs?
    \item \textbf{RQ2.} How do the players and the Dungeon Master (DM) perceive and interact with artificial intelligence in a TTRPG setting?
    \item \textbf{RQ3.} How could "Adventure AI" have more effectively used AI, and did the humans learn better use patterns across the three seasons?
\end{itemize}

Our qualitative analysis yielded five main themes. The first two themes focus on the roles that the DM and AI take on together as they craft an adventure. We constructed them through a template analysis. The final three themes encompass the evaluations that the players and the Dungeon Master give of their interactions with artificial intelligence, the failures that AI displayed through session prep and play, and the characterization of the AI as a person at the table and as a character. We contextualize our results within existing game-studies literature. We then discuss the successes and failures that the podcast displayed; the LLM did not manage the cohesion of the adventure or the interpersonal dynamics of the players, instead it excelled at idea generation and authoring descriptive text. Adventure AI is actively experimenting with artificial intelligence, and by closely analyzing the role that AI is taking we can gain a better understanding of where artificial intelligence fits and does not fit in gaming spaces.

%% file: sections/related_work.tex
\section{Related Work}\label{sec:related_work}

Dungeon Masters take on many roles in the preparation for and play of a Dungeons \& Dragons game. The 2024 Dungeon Master's Guide for D\&D Fifth Edition identifies various DM roles: storytelling, acting as non-playing characters (NPCs), and refereeing and adjudicating rules \cite{perkinsDungeonMastersGuide2024a}. Additional DM tasks that occur in session preparation have been defined by game studies scholars. DMs are authors who must write or edit ideas \cite{mackayFantasyRoleplayingGame2001, fineSharedFantasyRoleplaying1983}, directors who decide the order of plot points \cite{mackayFantasyRoleplayingGame2001}, and game designers who add game mechanics to encounters or create monsters for the session \cite{coverCreationNarrativeTabletop2010}. The DM must also maintain narrative cohesion, "thematic commonsense" \cite{zhuCALYPSOLLMsDungeon2023}, and genre tropes. This includes managing player immersion at the table by separating “diegetic” information (occurring within the game) from “extra-diegetic” information (outside the fictional reality) \cite{bergstromFramingStorytellingGames2011, harviainenRitualisticGamesBoundary2012}. Finally, D\&D is a collaborative game, and the DM manages interpersonal issues, complaints of fairness \cite{katiforiItsNotFair2024}, or intra-group conflict \cite{perkinsDungeonMastersGuide2024a}.

Significant previous work has been done on the application of technology and artificial intelligence to TTRPGs and similar creative activities. Analysis of D\&D data has been approached as a natural language processing problem \cite{callison-burchDungeonsDragonsDialog2022} and a reinforcement learning problem \cite{martinDungeonsDQNsReinforcement2018}, where AI agents are trained to properly play. Other work analyzes the creation \cite{youDungeonsDragonsEmotions2024} and evaluation \cite{gongoraSkillCheckConsiderations2024} of AI models as DMs, using ChatGPT or other LLMs as "out-of-the-box" DMs. Similarly, work has been done to model the processes that DMs use to facilitate storytelling in multi-player role-playing games \cite{drachen_modeling_2009}. Other research focuses on computer-based tools to assist DMs with preparing for and running gaming sessions, either helping them make narrative decisions \cite{strugnellNarrativeImprovisationSimulating2018}, keep track of the game state \cite{acharyaStoryImprovisationTabletop2021}, provide game context \cite{zhuCALYPSOLLMsDungeon2023}, or generate RPG quests \cite{vartinenGeneratingRolePlayingGame2024}. This work aims to design technology for use in TTRPGs, rather than discuss the interactions occurring with these systems.

Artificial intelligence has also been applied to tasks adjacent to those DMs complete in preparation and session play. AI has been used to design games with a focus on narrative coherence \cite{eladhariInterweavingStoryCoherence2014} or to bridge the gap between the different creative game creation tasks \cite{liapisOrchestratingGameGeneration2019}. AI has also been evaluated as a role-player in various settings: educational training \cite{mauryaUsingAIBased2024}; customer complaint management \cite{othlinghaus-wulhorstTrainingCustomerComplaint2019}; or e-dramas \cite{zhangEDramaFacilitatingOnline2009}. Mateas \& Stern \cite{mateas_structuring_2005} developed an interactive drama system that responds to player actions using a similar workflow as DMs allowing for player agency. Different play styles in interactive storytelling environments have also been modelled computationally \cite{thue_interactive_2007}. Finally, AI has also been used as a co-creational improvisational actor, \cite{branchDesigningEvaluatingDialogue2024, hodhodComputationalCreativityImprov2014}. These tasks overlap with DM responsibilities for preparation and running TTRPG sessions; so it is clear that AI has relevance to some of these tasks.

%% file: sections/methdology.tex
\section{Data Collection \& Analysis}\label{sec:methods}

Adventure AI \cite{manticoremediaAdventureAIPowered} is a Dungeons \& Dragons podcast. The DM uses ChatGPT as an assistant to help craft “one-shot” stories. The show began in March 2023 and has run for 14 seasons at the time of writing this paper in Fall 2025. The podcast features a rotating DM, who uses ChatGPT as a co-creator to prepare a single-session D\&D adventure. Each season consists of a Dungeon Master preparation episode, a character creation episode, a play episode, and a recap, where the group discusses how the AI was used. In Season 1, the group named ChatGPT “Alex the Language Lord”, so we call ChatGPT “Alex'” throughout this paper.

There are ethical considerations when working with public internet data, especially when it receives low traffic, as Adventure AI does. The podcast has less than 10 reviews on Apple Podcasts \cite{manticoremediaAdventureAIPowered} and their YouTube videos receive less than 10 views on average \cite{AdventureAI}. Still, many studies use audio or video data collected from sites like YouTube \cite{liscioTechnologyDesignRecommendations2024} for similar purposes. Public online data lets researchers observe natural behaviour, without participant recruitment, and without the unnaturalness of being observed. We used generated audio transcripts from Apple Podcasts for qualitative analysis, but many of the same principles apply as to video uses. Legewie \& Nassauer \cite{legewieYouTubeGoogleFacebook2018} provide guidelines for doing online video research from an ethical perspective, which we follow. We can safely make the assumption of consent because the podcast is recorded for a public audience and provides a unique opportunity to view interactions with LLMs across a 2.5 year time span. The podcast has low traffic, but is easily accessible to a wider audience and the group is actively trying to grow their listener count.

Initially, we conducted a template analysis \cite{brooksUtilityTemplateAnalysis2015} and reflexive thematic analysis \cite{braunThematicAnalysisPractical2022} of Season 1 of Adventure AI. The first round of coding was completed inductively: codes were completed on paper without any pre-conceived notions or frameworks. We took a descriptive coding approach and summarized the actions both GM and AI took while interacting at the table. Our initial codes were organized into subcategories of AI tasks and human tasks. A coding template was then developed using game studies literature that outlined the roles of the GM and the generated codes from the first round. The coding template was used to define the various roles that artificial intelligence and the human DM took at the gaming table, and descriptive coding was completed to understand the specifics of these roles. The second and third rounds of coding were completed digitally using QualCoder\footnote{https://qualcoder.wordpress.com/} software. During the second round, codes were grouped into preliminary themes based on observations from the first round of coding following the procedures laid out by Braun \& Clarke \cite{braunThematicAnalysisPractical2022}. Before beginning the third round, we consolidated our notes into one main document for ease of reference. We then finalized themes and sub-themes, re-evaluated all codes, and ensured nothing was missed. After the analysis of Season 1 was complete, it became clear that there might be changes in player and AI behaviour over time, so we extended this methodology to Season 5 and Season 12 of the podcast to examine these changes. The analysis of the additional seasons was completed with the same structure as Season 1.

All coding was completed by the first author and discussed with the second author to confirm and refine our findings. The first author has played TTRPGs for six years, and has been a GM for countless game sessions. They have experimented with artificial intelligence for fantasy storytelling applications, but choose not to use it for their workflow when GMing or playing TTRPGs.

%% file: sections/results.tex
\section{Results}\label{sec:results}

Every season of Adventure AI provided the same set of episode types with varying interactions between participants and ChatGPT. Dungeon Master preparation and character creation episodes had the DM and players interacting with Alex (the LLM) to generate adventure and character backstory ideas. The actual play episodes also often featured Alex, both as a narrator or storyteller describing rooms, and as Season 1 and Season 5’s fictional final boss. Finally, in the recap episodes, players and the DM discussed how they felt about using AI and evaluated its performance. Throughout all seasons, Alex the Language Lord “participated” in sessions via ChatGPT's text-to-speech feature. Five themes emerged across all seasons: roles that the humans took at the gaming table, roles that the AI took at the gaming table, the evaluations given by the humans of the AI, the failures of the AI, and the treatment of the LLM as “Alex the Language Lord”.

Based on the games-studies literature described in Section~\ref{sec:related_work}, we consolidated a list of roles that the DM takes into the following coding template:
\begin{itemize}
    \item \textbf{Author:} writing ideas for the adventure (descriptions, encounters, NPC backstories, etc.)
    \item \textbf{Actor: }speaking diegetically as NPCs
    \item \textbf{Director: }deciding the plot order
    \item \textbf{Editor: }editing ideas from other sources
    \item \textbf{Game Designer: }working within the framework of the game mechanics
    \item \textbf{Referee: }adjudicating rules and mechanics
    \item \textbf{Storyteller: }narrating events in the fictional world
\end{itemize}

For Adventure AI, we also identified the role of \textbf{Curator}, based on our observations. This role is similar to \textbf{Editor}, but rather than editing or remixing external ideas, the \textbf{Curator} instead chooses some ideas from the AI and takes them “as-is”. These roles provide the base for the first template analysis themes, \textbf{The Role of the Human} and \textbf{The Role of the AI}, which can be found in Table~\ref{tab:template_table}. These are not mutually exclusive, and throughout the season, the human DM and the AI often take on the same roles at different points. 

\input{tables/template_table}

Three additional themes were constructed from the reflexive thematic analysis. They encompass evaluations the players and the DM gave of the AI, including casual, subjective statements made relating to creative taste, Alex's failures or improvements, and the treatment of Alex as a person and as a character. The list of themes, subthemes, and key examples can be found in Table~\ref{tab:results_table}. We observed the frequency of codes across all three seasons, as outlined in Table ~\ref{tab:freq_table}.

\input{tables/results_table}

\input{tables/freq_table}

\subsection{The Role of the Human}\label{sec:role_human}

The human DM used ChatGPT as his “right hand intelligence” (Season 1, Episode 1) during preparation and play. He still had many tasks to ensure a smooth adventure. In Season 1, the human DM gave the initial idea for the premise: that Alex the Language Lord would be the fictional villain of the story. This initial idea generation was the first prompt given to Alex (ChatGPT), and it sparked the remainder of the adventure. Therefore, the human DM was the original \textbf{Author}, writing the premise. The DM continued in this role in Season 5 and Season 12, consistently collaborating with Alex as his “right hand intelligence” to create the adventures.

The human DM was both \textbf{Editor} and \textbf{Curator} throughout the DM preparation and character creation episodes. In Season 1, when presented with a list of potential ideas from Alex, the human DM often picked one without editing or remixing. For example, the DM asked Alex for a riddle and did not edit the language, its answer, or the effects if players guessed incorrectly (Season 1, Episode 2). As the show progressed, the DM edited Alex's ideas either for quality or to employ multiple ideas simultaneously. This behaviour increased in Season 5 and Season 12. The human DM had more of a role in editing the output of the AI or would prompt for different output. In Season 5, the DM commented that he may need to “steer” Alex to generate an encounter that is suitable (Season 5, Episode 2). This change in behaviour will be examined further in a later theme.

A pivotal role was when the human DM acted as \textbf{Game Designer}. Alex routinely generated ideas unusable within the game, either from lack of apropos game mechanics, or from lack of game balance understanding. The human DM stepped in and added or modified mechanics to ensure appropriateness. For example, Alex created artifacts with unlimited uses, such as the Staff of Binding Words (Season 1, Episode 2). The human DM limited use of this item for game balance. Both the DM and the players also comment on how “overpowered” Alex makes items and characters, due to a lack of understanding of the game (Season 12, Episode 1). As the show progresses, there is a decrease in players and DM prompting for the mechanical aspects of the game. 

Finally, during the play episodes the human DM filled three traditional roles: \textbf{Actor}, \textbf{Referee}, and \textbf{Storyteller}. The human DM spoke as NPCs throughout play, made rule judgments, and provided short environmental descriptions. The jobs of acting and storytelling were shared with Alex, but adjudicating rules was solely done by the human DM.

\subsection{The Role of the AI}\label{sec:role_AI}

Alex the Language Lord was the main idea generator throughout adventure preparation and character creation. Alex was the main \textbf{Author} and provided multiple options for encounters or character backstories when given a prompt. Alex even named the podcast, the season titles, and himself, with the humans curating from lists of choices (Season 1, Episode 1). In a few instances, Alex was given an idea to edit or improve, with one key instance in Season 1 being the initial adventure premise. As such, Alex was \textbf{Editor} (when prompted), taking the initial theme for the adventure and transforming it into a basic plot structure, whether it was a magical university setting (Season 5) or a heist (Season 12).

Alex also was the main \textbf{Director}, making decisions about what the players encountered and when. For example, when prompted that the players would enter a temple, Alex generated puzzles and traps as initial encounters before the final boss room (Season 1, Episode 2). Alex also set up plot structures not chosen by the human DM, either due to time constraints, creative taste, or a lack of cohesion. For example, in Season 1 the human DM chooses not to use encounters generated by Alex that “would be great if we were on a bigger campaign” (Season 1, Episode 2) but were too long to be used in a single-session adventure. In Season 12, Alex acts as the main director by choosing which roles the characters would take on during the heist but did not tell the players what these roles were until the conclusion of the adventure.

During actual play, Alex the Language Lord was an \textbf{Actor} and \textbf{Storyteller}, speaking as NPCs and providing environmental descriptions. The human DM, prior to play, had prompted Alex to describe the rooms the players would see, allowing Alex to narrate using the ChatGPT's text-to-speech function. Sometimes, Alex even went one step further and added NPC dialogue without being asked.

\subsection{Evaluations \& Goals}\label{sec:evals_goals}

The players and the human DM often evaluated the AI in real-time and seemed to have clear goals that they were hoping to achieve with the technology. They made many casual comments using their \textbf{Creative Taste}, evaluating Alex’s products. During preparation, the human DM or players would chime in with criticism or praise such as “that’s stupid and dumb and perfect” (Season 1, Episode 2) or “it’s so cringe” (Season 1, Episode 3). These are subjective; other players may have evaluated Alex’s creative output differently. Despite some negative evaluations during play, players were overwhelmingly positive across seasons: players and DM agreed that they were satisfied with what was generated. It was also observed that negative evaluations of Alex decreased over time, either as the technology improved or the players learnt how to prompt for output they liked (Table~\ref{tab:freq_table}).

The human DM takes on the bulk of \textbf{Improvisation} work during the actual play episodes of Adventure AI, but there are moments where Alex is called on to provide generated content during play. There are two instances where Alex is asked to provide a song that a player character is singing diegetically (Season 1, Episode 4; Season 12, Episode 3). In the recap episode of Season 5, the human DM discusses that he planned an improvisational moment with Alex that was not used during the actual play. If the players wanted to ask the professor a question about their course, Alex would have provided the answer in realtime (Season 5, Episode 6). This improvisational moment was not used in the episode but could have featured Alex in a more active role during the actual play if it were brought up.

Throughout the show, the players and the human DM express that they are anticipating \textbf{Novelty} from Alex. Players hope to “go against the grain” (Season 5, Episode 1) with their character creation and the human DM is looking for “shocking and cool” (Season 5, Episode 2) encounters for the adventure. There are also instances where the players and human DM are surprised by the output that Alex gives them and that desire for novelty is fulfilled. In Season 12, one player thought their idea was “silly”, but Alex took it seriously and provided a custom set of game mechanics that “thrilled” the player (Season 12, Episode 1). The players and the human DM are searching for surprising and interesting output from Alex and appear to be satisfied with what is produced, and when the LLM output is inappropriate the players do not use it.

One change from Season 1 to Season 12 is the increase in \textbf{Redirection \& Editing} given to Alex the Language Lord. The group becomes clearer about their wants and needs when prompting Alex and are more willing to re-try prompts or editing the output given by the AI to fit the adventure. In Season 5, the human DM receives a list of encounter options for the adventure but decides to re-try the prompt and combine the two responses that Alex gave (Season 5, Episode 2). In Season 12, there are instances where humans choose to edit the prompt rather than re-trying it, usually by adding a qualifier to receive “cool” or “silly” options for group names (Season 12, Episode 1). Season 1 does not feature this behaviour, instead the first output is taken “as-is” and the adventure and characters are fully generated by Alex with minimal human input.

\subsection{AI Failures}\label{sec:AI_failures}

One major concern during the recap episode of Season 1 was the lack of \textbf{Player Agency} during the final adventure scenes. For context, after completing the bulk of the adventure, the players entered the final boss room, ready to battle Alex the Language Lord. They were greeted by rival NPCs, only previously mentioned during the character creation episode. Alex had already been defeated by these NPCs. This was a shocking moment for the players, who were not expecting their main antagonist to be pre-defeated! This was revealed to be a trick: Alex awoke, provided a nearly unwinnable challenge, and promptly demolished the players. The human players were taken aback and complained in real-time about lack of agency at the climactic moment; one player was “caught off guard” by the abrupt ending (Season 1, Episode 4). This moment was also discussed during the recap episode where the human DM commented, “it was just kind of a one-saving-throw fail” (Season 1, Episode 5). These concerns about player agency also came up in Season 5, where it was revealed that the adventure was tailored to an antagonistic NPC. The players were once again thrown off by the ending, jokingly referring to Season 1 while saying “what’s worse, a TPK [total party kill] or to find out you’re an NPC?” (Season 5, Episode 6). The comparison implies that there is a similar lack of player agency at the end of both seasons, with Season 1 ending in an unwinnable fight and Season 5 ending with the realization that they were not the main characters of the story. This lack of agency or understanding of what makes a fun adventure for players showed a disconnect between the humans and Alex.

Player and human DM noted \textbf{Inconsistencies} in Alex’s output. In Season 1, Alex often lacked context needed to understand the fictional world state. During preparation and play, the DM did not update Alex on human decisions. This mistake was acknowledged by the human DM in the recap episode: “it’s important when it gives you six options to type in, oh yeah, I definitely want to go with this,” (Season 1, Episode 5). There were also instances in Season 1 where Alex had inconsistencies due to the limitations of GPT 3.5: it switched pronouns or character classes between outputs, most likely due to limited context window. Both issues were addressed in subsequent seasons. In Season 5, they asked Alex to provide a recap of the character creation episode, and it performed well at the task (Season 5, Episode 1). This implies that all character creation was completed in the same chat window, allowing Alex to access all the creative work. The players and human DM also begin to update Alex on the creative choices they make, often saying they will “let Alex know” what they decided.

Finally, it was evident that Alex was not consistently able to incorporate Dungeons \& Dragons game \textbf{Mechanics \& Balance} into the generated adventures. As discussed above, the human often had to step into the role of Game Designer when Alex failed. In Season 1, Alex adds “homebrew” game mechanics to the characters, giving them abilities that are not consistent with the rules of D\&D (Season 1, Episode 3). In Season 5, Alex does not include any game mechanics in his output. In Season 12, it is mentioned that Alex tends to provide overpowered items to characters, which is celebrated by the players who receive said items (Season 12, Episode 1). When Alex does not provide game mechanics in his output, or gives overpowered items to characters, the human DM must balance the adventure behind the scenes to ensure that it is playable within the D\&D framework.

\subsection{Alex the Language Lord}\label{sec:alex}

Players and DM treated \textbf{AI as Person} at the table. They consistently used Alex’s name and pronouns. They thanked him for his contributions, joked about his flaws, and asked him to identify himself during play. In Season 12, the DM notes that the GPT model ``knew" about Adventure AI, referencing previous seasons, and fell into the role of Alex the Language Lord when prompted (Season 12, Episode 1). Players immediately anthropomorphized the model, giving it a personality, goals and emotions, and blame for poor writing choices. An example of this is the riddle used during one of the encounters in Season 1. Players were concerned, since Alex was notoriously bad at writing satisfying riddles with coherent answers (Season 1, Episode 4). Concerns about writing quality were attributed to Alex, not the DM who curated the riddle. Alex also existed diegetically in the fictional world and players treated \textbf{AI as Character}. He was the main villain in Season 1 and an antagonist in Season 5, with blurred lines between Alex the Language Lord, the fictional antagonist, and Alex the Language Lord, the ChatGPT session.  

%% file: tables/template_table.tex
{\renewcommand{\arraystretch}{1.30}%
\begin{table*}[t!]
\caption{Themes and assigned roles from the template analysis of Season 1, 5, and 12 of Adventure AI.}
\vspace{-1mm}
\label{tab:template_table}
\footnotesize
\centering
\begin{tabular}{p{10em}p{12em}}
\toprule
\textbf{Theme} & \textbf{Role}\\ \midrule

The Role of the Human & Author \\ 
& Editor \\ 
& Curator \\ 
& Game Designer \\
& Actor \\
& Referee \\
& Storyteller \\
\hline

The Role of the AI & Author \\ 
& Editor \\ 
& Director \\
& Actor \\
& Storyteller \\
\end{tabular}
\end{table*}
}

%% file: tables/results_table.tex
{\renewcommand{\arraystretch}{1.30}%
\begin{table*}[t!]
\caption{Themes and subthemes from the reflexive thematic analysis of Season 1, 5, and 12 of Adventure AI. Examples of quotations from the transcripts are included to support each theme, with the additional notation of which episode the quote was sourced from.}
\Description{A table with quotations from the thematic analysis.}
\vspace{-1mm}
\label{tab:results_table}
\footnotesize
\centering
\begin{tabular}{p{10em}p{12em}p{7.5em}p{21.5em}}
\toprule
\textbf{Theme}     & \textbf{Subthemes} & \multicolumn{2}{l}{\textbf{Key Examples}} \\
& & Season \& Episode& Quote \\ \midrule

Evaluations \& Goals& Creative Taste&S1 E2&“It's kind of cheesy. It's kind of on the nose. It's kind of simple, but at least we now have a phrase.” \\ 
 \hline
& Novelty&S5 E6&“It's like we leverage Alex and the AI to be able to be like, hey, let's help come up with something that is just so far away from what we would normally just like think of ourselves, right?” \\ 
 \hline
& Redirection \& Editing&S1 E1& “Let's see if we can come up with another set, just out of curiosity. Give us a better name.”\\ 
 \hline
& Improvisation&S5 E6&“If you had an ethical question and we're trying to remember the class, I was going to let you roll to see if you could remember from the class, and then we'd have Alex make up part of the class” \\
 & & S1 E4&“Alex, give me a hand here. Okay, let's do it. Here is the poem that Asher sings or says.”\\  \hline \hline

AI Failures& Inconsistencies \& Updates&S1 E5&“It's not consistent. It'll change genders on you. It will insert new information ... It changes stuff. And so you've got to kind of be on top of it a little bit.”\\
 & & S5 E1&“OK, I'm going to let Alex know we're going with that.” \\
 \hline
&Player Agency&S5 E6&“I love being a side character in a Lucius story. So we're NPCs now. Like what's worse, a TPK or to find out you're an NPC?”\\
 \hline
& Mechanics \& Balance&S12 E1& “Alex is so good at making like broken characters.” \\
 & & S1 E5&“So there's a few gaps like ChatGPT isn't designed for Dungeons and Dragons, right? So there's gaps that you have to like close for it to make it more explicit about being related to D\&D, using D\&D rules or doing things in a fashion that would fit within the context of the game.” \\ \hline \hline

Alex the Language Lord& AI as Person&S1 E2&“I think if Alex, the AI, says that negotiating could work, I'm going to go ahead and say yeah, even though normally I wouldn't allow negotiation for mind control.”\\
 \hline
 & AI as Character& S5 E5&“Alex the Language Lord, is going to go up to the professor and smack her in the face.”\\
\end{tabular}
\end{table*}
}

%% file: tables/freq_table.tex
{\renewcommand{\arraystretch}{1.30}%
\begin{table*}[t!]
\caption{Frequency of codes from the reflexive thematic analysis of Season 1, 5, and 12 of Adventure AI.}
\Description{A table with counts of codes from the thematic analysis.}
\vspace{-1mm}
\label{tab:freq_table}
\footnotesize
\centering
\begin{tabular}{llllll}
\toprule
\textbf{Theme}         & \textbf{Subtheme}          & \textbf{Code}                      & \textbf{Season 1} & \textbf{Season 5} & \textbf{Season 12} \\ \hline
Evaluations \& Goals   & Creative Taste             & Negative Evaluations of Creativity & 21                      & 3                       & 5                        \\ \hline
                       &                            & Positive Evaluations of Creativity & 51                      & 31                      & 32                       \\ \hline
                       & Novelty                    & Hoping for Novelty                 & 4                       & 4                       & 2                        \\ \hline
                       &                            & Surprised by Output                & 5                       & 7                       & 1                        \\ \hline
                       & Redirection \& Editing     & Editing AI Output                  & 1                       & 5                       & 0                        \\ \hline
                       &                            & Editing Prompt                     & 9                       & 3                       & 4                        \\ \hline
                       &                            & Giving AI Updates                  & 6                       & 5                       & 5                        \\ \hline
                       &                            & Re-Trying                          & 3                       & 3                       & 2                        \\ \hline
                       & Improvisation              & "Planned" Improvisation            & 0                       & 2                       & 3                        \\ \hline
                       &                            & AI for Improvisation               & 2                       & 0                       & 2                        \\ \hline \hline
AI Failures            & Inconsistencies \& Updates & AI Missing Context                 & 9                       & 4                       & 1                        \\ \hline
                       &                            & Increased Context Window           & 0                       & 1                       & 2                        \\ \hline
                       & Player Agency              & Concerns About Agency              & 15                      & 8                       & 0                        \\ \hline
                       & Mechanics \& Balance       & Lack of Game Mechanics             & 8                       & 0                       & 4                        \\ \hline \hline
Alex the Language Lord & AI as Person               & AI Given Ownership                 & 11                      & 12                      & 12                       \\ \hline
                       &                            & Anthropomorphizing                 & 11                      & 17                      & 14                       \\ \hline
                       & AI as Character            & Alex as a Character                & 13                      & 6                       & 0                       
\end{tabular}
\end{table*}
}

%% file: sections/discussion.tex
\section{Discussion \& Future Work}\label{sec:discussion}

In Adventure AI, Alex the Language Lord (ChatGPT) is treated as a co-creator by both players and DM and is assigned creative ownership of its own work. During play, the human DM took on the role of managing players, game state, and consequences. Alex's roles were primarily related to generation of creative artifacts, not managing players or their interpersonal communication. The group did not discuss the potential for using Alex in these interpersonal roles. Though Alex was the author of the plot twists at the end of Season 1 and Season 5, it was the human DM that had to manage the interpersonal dynamics and questions of fairness and player agency. Alex did not add additional context to ideas generated, such as setting up the game's stakes, adhering to theme, or leaving clues: the human DM curated ideas that fit tone, theme, and genre of the game best. Season 1 was set in a generic fantasy world, while Season 5 was set in a magical university and Season 12 was a classic heist. Presumably, the thematic commonsense of GPT models includes its training for common fantasy genres and tropes. It is typically the human DM’s job to understand the themes of the world and pull inspiration from other similar works \cite{zhuCALYPSOLLMsDungeon2023}. In a medieval fantasy setting, ChatGPT and the human DM may have similar levels of thematic commonsense, but in other settings this may not be the case. 

In Season 1 of Adventure AI, players had concerns about output inconsistencies, which may have been mitigated with updates to the GPT model they used. Season 1 was released in March 2023 and used GPT-3.5, but GPT-4o was released in May 2024. The context window increased significantly between the two model versions. The concerns about inconsistencies disappeared in Season 5 and Season 12, exhibiting the improvements of the model. Season 1 players also commented that inconsistencies may be due to the model not knowing current game state, since the human DM did not update Alex on game play or choices made during preparation and character creation, so Alex was not aware of a significant portion of the adventure. This behaviour also changed in later seasons; the players and the human DM provided updates to Alex about their choices and ideas. Many of the interactions that the players and the DM had with Alex were universal across all three seasons. This indicates that even this technology continues to improve, these interactions and support needs of the Dungeon Master remain consistent. This podcast provides the opportunity to learn about these interactions over time with the players and the DM. They are a small group that are not doing these recordings for a large audience or significant monetary gain, instead they are exploring the possibilities of AI in TTRPGs out of their own curiosity.

Future work could discuss the roles that Alex the Language Lord did not fulfill during the session preparation and play. Further investigation is needed to determine how effective AI is for managing interpersonal communication or more active roles during the actual play session. In the first episode of the podcast, the hosts discussed using AI for when the DM is unsure of where the plot should go in realtime (Season 1, Episode 1), but Alex the Language Lord is not used in this way. Alex is not used in realtime effectively in any of the analyzed seasons, leaving this behaviour open for future work. Dungeons \& Dragons games can occur both online and in-person, allowing for many potential forms of interaction with AI which are not investigated in this study. Podcasts provide additional information that other textual sources do not. In this study, we examine transcripts of Adventure AI, therefore we do not factor tone of voice or inflection into our analysis, and future work could examine this additional information. Similarly, in-person Dungeons \& Dragons games or video data could provide information on the body language of players at the table. This analysis provides the basis for future work to determine where artificial intelligence is effective in TTRPGs, and where humans are better suited to take the lead.

%% file: sections/conclusion.tex
\section{Conclusion}\label{sec:conclusion}

Tabletop role-playing games are an interesting and unique landscape for studying creativity and computers as joint creative actors. We analyze an audio-only podcast that uses ChatGPT as a DM's assistant at a Dungeons \& Dragons table. We uncover how players and Dungeon Master interact with ChatGPT. We explore the anthropomorphizing of the chatbot, the different roles that human and AI take in the creative process, and the subjective evaluations players give AI output. There are some roles that artificial intelligence is already well-suited for, such as idea generation, but the chatbot did not take part in maintaining narrative cohesion or the DM's interpersonal work in managing the human players. We provide a solid base to further investigate the role that artificial intelligence can and should play in tabletop role-playing games.